\documentclass[aps,prl,showpacs,psfig,twocolumn]{revtex4}
\usepackage{amssymb}
\usepackage{amsmath}
\usepackage{graphicx}
\usepackage{dcolumn}
\usepackage{bm}
\usepackage{psfig}

\input{tcilatex}

\begin{document}

\title{Spin-Polarized Transport in Ferromagnet-Marginal Fermi Liquid Systems}
\author{Hai-Feng Mu, Gang Su$^{\ast }$, Qing-Rong Zheng and Biao Jin}
\affiliation{College of Physical Sciences, Graduate School of the Chinese Academy of
Sciences, P.O. Box 3908, Beijing 100039, China}

\begin{abstract}
Spin-polarized transport through a marginal Fermi liquid (MFL) which is
connected to two noncollinear ferromagnets via tunnel junctions is discussed
in terms of the nonequilibrium Green function approach. It is found that the
current-voltage characteristics deviate obviously from the ohmic behavior,
and the tunnel current increases slightly with temperature, in contrast to
those of the system with a Fermi liquid. The tunnel magnetoresistance (TMR)
is observed to decay exponentially with increasing the bias voltage, and to
decrease slowly with increasing temperature. With increasing the coupling
constant of the MFL, the current is shown to increase linearly, while the
TMR is found to decay slowly. The spin-valve effect is observed.
\end{abstract}

\maketitle

Spin-polarized transport in magnetic hybrid nanostructures has been an
active subject under investigation in last decades, which is mainly
motivated by potential applications in information technology. A new field
coined as spintronics is thus emerging (for review, see e.g. Refs. \cite%
{Prinz,J.S.Moodera,Wolf,das,Igor}). The well-known character in spintronics
is that the current flowing through the structures depends sensitively on
the relative orientation of the magnetization directions due to the
spin-dependent scattering of conduction electrons. Among others, the
magnetic tunnel junction (MTJ) is an important family of spintronic devices%
\cite{maekawa,su}. For these structures, Julli\`{e}re \cite{Julliere} was
the first to observe the tunnel magnetoresistance (TMR) of $14\%$ in
Fe-Ge-Co junctions at $4.2$ $K$. In 1995, Moodera \textit{et al}. \cite%
{Moodera} made a breakthrough by observing reproducibly a large TMR as high
as $24\%$ at $4.2K$ and $11\%$ at $295K$. Recently, clear spin-valve signals
at $4.2K$ as well as at room temperature have been observed in
ferromagnet-normal metal-ferromagnet (FM-N-FM) all-metal structures\cite%
{jedema}. Earlier theories on the spin-dependent transport in FM-N-FM
junctions\cite{brataas} are based on the Fermi liquid theory, where
interactions between electrons in the normal metal are treated on a
mean-field level. There has been recent studies on the spin transport in
FM-Luttinger liquid-FM tunnel junctions where the interactions between
electrons are taken into account, and applied directly to carbon nanotubes%
\cite{balents,mehrez}, but they are primarily aimed at one-dimensional
interacting quantum wires. Besides, spin-polarized transport through an
interacting quantum dot that is described by the Anderson model has also
gained much attention\cite{serg}. On the other hand, there appear intriguing
experimental and theoretical works on the spin-polarized transport in
FM-high Tc superconductor tunnel junctions recently (e.g. Refs.\cite{vasko}%
). It is thought that the anomalous normal state properties of high Tc
cuprates in the optimally doped regime can be well described by the marginal
Fermi liquid (MFL)\cite{Varma}, where the interactions between electrons in
the cuprates are phenomenologically included in a one-particle self-energy
due to exchange of charge and spin fluctuations. Therefore, the study on the
spin-dependent transport in FM-MFL-FM tunnel junctions would be interesting,
as it would be useful for understanding the transport properties of FM-high
Tc cuprate junctions in the normal state.

In this paper, by using Keldysh's nonequilibrium Green function formalism,
the spin-dependent transport in FM-MFL-FM tunnel junctions is investigated.
It is observed that the current-voltage characteristics in this spintronic
structure show non-ohmic behaviors, and the tunnel current increases slowly
with temperature, which are in contrast to those of the structure with a
Fermi liquid, showing that the interactions between electrons in the normal
metal have remarkable effects on the transport properties. The TMR is found
to decay exponentially with increasing the magnitude of bias voltage, and to
decrease slowly with increasing temperature. With increasing the coupling
constant $\lambda $ of the MFL, the current is shown to increase linearly,
while the TMR is seen to decay slowly, implying that the interactions of
electrons tend to suppress the TMR. In addition, the spin-valve effect is
observed.

Let us consider a MTJ in which the two FM electrodes, connected with the
bias voltage $V/2$ and $-V/2$, respectively, are separated by a normal metal
which is described by the MFL, as schematically depicted in Fig. 1. The
molecular field $\mathbf{h}_{L}$ in the left (L) FM is assumed to be
parallel to the $z$ axis, while the molecular field $\mathbf{h}_{R\text{ }}$%
in the right (R) FM is parallel to the $z^{\prime }$ axis which deviates the 
$z$ axis by a relative angle $\theta $. $T_{k\alpha q}$ ($\alpha =L,R$)
stand for the elements of the tunneling matrix between the $\alpha $
electrode and the central region. The tunnel current flows along the $x$
axis and perpendicular to the junction plane. In the central region, the
interactions between conduction electrons are supposed to be described
phenomenologically by a retarded one-particle self-energy due to the
exchange of charge and spin fluctuations\cite{Varma}: 
\begin{equation}
\Sigma (\varepsilon )=\lambda \lbrack \varepsilon \ln \frac{x}{E_{c}}-i\frac{%
\pi }{2}x],  \label{self-e}
\end{equation}%
where $x=\max (|\varepsilon |,k_{B}T)$, $E_{c}$ is a cut-off energy and $%
\lambda $ is a coupling constant. When $\lambda =0$, the MFL junction
recovers the conventional Fermi liquid. For simplicity, the spin-orbital
coupling in the MFL will be ignored. It is worthy of noting that the exact
Hamiltonian of the MFL is not yet available. However, since the
single-particle Green function is explicitly written down, the concrete form
of the microscopic Hamiltonian is irrelevant. In the calculations, we just
need to adopt a formal Hamiltonian such as a Fermi liquid with the electron
operators understood as those of quasi-particles. Because the final results
are all expressed by Green functions, we only need to use the MFL Green
functions to replace the quasi-particle Green functions. 
\begin{figure}[tb]
\begin{center}
\leavevmode\includegraphics[width=0.75\linewidth,clip]{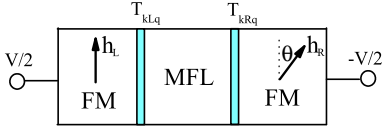}
\end{center}
\caption{(Color online) Schematic illustration of the double tunnel junction
consisting of two ferromagnets (FM) and a marginal Fermi liquid (MFL)
separated by insulating films, where $T_{k\protect\alpha q}$ ($\protect%
\alpha =L,R$) stand for the elements of coupling matrix between the $\protect%
\alpha $ electrode and the central region, and both magnetizations are
aligned by a relative angle $\protect\theta $.}
\label{fig1}
\end{figure}

By means of the nonequilibrium Green function, the tunnel current through
the left electrode can be obtained by 
\begin{equation}
I_{L}(V)=e\langle \dot{N}_{L}\rangle =-\frac{2e}{\hbar }\Re e\sum_{\mathbf{kq%
}\sigma }T_{\mathbf{k}L\mathbf{q}}G_{\mathbf{q}\sigma L\mathbf{k}\sigma
}^{<}(t,t),  \label{current}
\end{equation}%
where $N_{L}$ is the occupation number of electrons in the left electrode, $%
G_{\mathbf{q}\sigma ^{\prime }L\mathbf{k}\sigma }^{<}(t,t^{\prime
})=i\langle a_{\mathbf{k}\sigma }^{\dag }(t^{\prime })c_{\mathbf{q}\sigma
^{\prime }}(t)\rangle $ is the lesser Green function, $a_{\mathbf{k}\sigma }$
and $c_{\mathbf{k}\sigma }$ are annihilation operators of electrons with
momentum $\mathbf{k}$ and spin $\sigma $ $(=\pm 1)$ in the left and central
region, respectively. In order to get the lesser Green function, we define a
time-ordered Green function $G_{\mathbf{q}\sigma ^{\prime }L\mathbf{k}\sigma
}^{t}(t,t^{\prime })=-i\langle T\{a_{\mathbf{k}\sigma }^{\dag }(t^{\prime
})c_{\mathbf{q}\sigma ^{\prime }}(t)\}\rangle $. In terms of the equation of
motion, we have 
\begin{equation*}
G_{\mathbf{q}\sigma ^{\prime }L\mathbf{k}\sigma }^{t}(t-t^{\prime })=%
\underset{\mathbf{q}^{\prime }}{\sum }\int G_{\mathbf{q}\sigma ^{\prime }%
\mathbf{q}^{\prime }\sigma }^{t}(t-t_{1})T_{\mathbf{k}L\mathbf{q}}g_{\mathbf{%
k}L\sigma }^{t}(t_{1}-t^{\prime })dt_{1},
\end{equation*}%
where $g_{\mathbf{k}L\sigma }^{t}(t)=(i\hbar \frac{\partial }{\partial t}%
-\varepsilon _{\mathbf{k}L\sigma })^{-1}$ with $\varepsilon _{\mathbf{k}%
L\sigma }=\varepsilon _{L}(\mathbf{k})-(eV/2)-\sigma M_{L}$, $\varepsilon
_{L}(\mathbf{k})$ the single-particle dispersion in the left electrode and $%
M_{L}=g\mu _{B}h_{L}/2$ ($g$: Land\'{e} factor, $\mu _{B}$: Bohr magneton),
and $G_{\mathbf{q}\sigma ^{\prime }\mathbf{q}^{\prime }\sigma
}^{t}(t-t^{\prime })$ is the time-ordered Green function in the central
region. By applying Langrenth theorem \cite{Haug} and Fourier transform, one
may obtain formally%
\begin{eqnarray}
G_{\mathbf{q}\sigma ^{\prime }L\mathbf{k}\sigma }^{<}(\varepsilon ) &=&%
\underset{\mathbf{q}^{\prime }}{\sum }T_{kLq^{\prime }}[G_{\mathbf{q}\sigma
^{\prime }\mathbf{q}^{\prime }\sigma }^{r}(\varepsilon )g_{\mathbf{k}L\sigma
}^{<}(\varepsilon )  \notag \\
&&+G_{\mathbf{q}\sigma ^{\prime }\mathbf{q}^{\prime }\sigma
}^{<}(\varepsilon )g_{\mathbf{k}L\sigma }^{a}(\varepsilon )],  \label{lesser}
\end{eqnarray}%
where $G_{\mathbf{q}\sigma ^{\prime }\mathbf{q}^{\prime }\sigma
}^{r}(\varepsilon )$\ is the Fourier transform of the retarded Green
function of electrons in the MFL of the central region, and $G_{\mathbf{q}%
\sigma ^{\prime }\mathbf{q}^{\prime }\sigma }^{<}(\varepsilon )$\ is the
corresponding lesser Green function, $g_{\mathbf{k}L\sigma }^{<}(\varepsilon
)$ and $g_{\mathbf{k}L\sigma }^{a}(\varepsilon )$ are the lesser and
advanced Green functions for the uncoupled electrons in the left electrode.
By defining $\Gamma _{\alpha }(\varepsilon )_{\mathbf{q}^{\prime }\sigma 
\mathbf{q}\sigma ^{\prime }}=2\pi D(\varepsilon )T_{\mathbf{k}\alpha \mathbf{%
q}}T_{\mathbf{k}\alpha \mathbf{q}^{\prime }}\delta _{\sigma \sigma ^{\prime
}}$ with $D(\varepsilon )$ the density of states (DOS) in the $\alpha $
electrode and using the Fourier transform, after a tedious but direct
derivation, Eq. (\ref{current}) can be rewritten as%
\begin{eqnarray}
I_{L}(V) &=&-\frac{ie}{\hbar }\int \frac{d\varepsilon }{2\pi }Tr\{\Gamma
_{L}(\varepsilon +\frac{eV}{2}+\sigma M_{L})  \notag \\
&&\times \lbrack f_{L}(\varepsilon )(G^{r}(\varepsilon )-G^{a}(\varepsilon
))+G^{<}(\varepsilon )]\},  \label{current-1}
\end{eqnarray}%
where $f_{\alpha }(\varepsilon )$ is the Fermi function of the $\alpha $
electrode, and $Tr$ is the trace over the momentum and spin space. Note that
in Eq. (\ref{current-1}) all Green functions, $G^{r,a,<}(\varepsilon )$, are
for electrons in the MFL of the central region, where $G^{r,a}(\varepsilon )$
are known with the presumed self-energy $\Sigma (\varepsilon )$ in the MFL
[Eq. (\ref{self-e})], say, $G^{r}(\varepsilon )=[\varepsilon -\varepsilon
_{k}-\Sigma _{0}^{r}-\Sigma (\varepsilon )+i\eta ]^{-1}$, where $\Sigma
_{0}^{r}$, $\Sigma (\varepsilon )$ denote the coupling of MFL to the two
ferromagnets and the retarded self-energy of the MFL, respectively, while $%
G^{<}(\varepsilon )$ is unknown and needs to be obtained.

To get the lesser Green function $G^{<}(\varepsilon )$ of the central
region, we invoke Ng's ansatz\cite{Ng}: $\Sigma ^{<}=\Sigma _{0}^{<}B$,
where $\Sigma _{0}^{<}(\varepsilon )=i[f_{L}(\varepsilon )\Gamma
_{L}(\varepsilon +\frac{eV}{2}+\sigma M_{L})+f_{R}(\varepsilon )R\Gamma
_{R}(\varepsilon -\frac{eV}{2}+\sigma M_{R})R^{\dag }]$, $B=(\Sigma
_{0}^{r}-\Sigma _{0}^{a})^{-1}(\Sigma ^{r}-\Sigma ^{a})$, $\Sigma
_{0}^{r}(\varepsilon )-\Sigma _{0}^{a}(\varepsilon )=-i[\Gamma
_{L}(\varepsilon +\frac{eV}{2}+\sigma M_{L})+R\Gamma _{R}(\varepsilon -\frac{%
eV}{2}+\sigma M_{R})R^{\dag }]$, $\Sigma ^{r}(\varepsilon )-\Sigma
^{a}(\varepsilon )=\Sigma _{0}^{r}(\varepsilon )-\Sigma _{0}^{a}(\varepsilon
)-i\lambda \pi x$, with $R=\left( 
\begin{array}{cc}
\cos \frac{\theta }{2} & -\sin \frac{\theta }{2} \\ 
\sin \frac{\theta }{2} & \cos \frac{\theta }{2}%
\end{array}%
\right) $ the rotation matrix, and $M_{R}=g\mu _{B}h_{R}/2$. Under this
presumption, one may find eventually that Eq. (\ref{current-1}) becomes 
\begin{eqnarray}
I_{L}(V) &=&\frac{e}{\hbar }\int \frac{d\varepsilon }{2\pi }%
Tr\{(f_{R}-f_{L})\Gamma _{L}(\varepsilon +\frac{eV}{2}+\sigma M_{L})  \notag
\\
&&\times G^{r}(\varepsilon )R\Gamma _{R}(\varepsilon -\frac{eV}{2}+\sigma
M_{R})R^{\dag }BG^{a}(\varepsilon )\}.  \label{current-2}
\end{eqnarray}

The TMR ratio can be defined according to the current as usual\cite%
{F.M.Souza}: 
\begin{equation}
TMR=\frac{I(\theta =0)-I(\theta =\pi )}{I(\theta =0)}.  \label{TMR}
\end{equation}%
When the magnetizations of the two FMs are noncollinearly arranged, the TMR
ratio can be described by 
\begin{equation}
TMR(\theta )=\frac{I(0)-I(\theta )}{I(0)}.  \label{TMR-theta}
\end{equation}%
Obviously, $TMR(\pi )=TMR$, and $TMR(0)=0$.

In the following, for the sake of simplicity for numerical calculations, and
considering that the electrons near the Fermi level in metals are dominant
in the tunneling process, we may suppose $\Gamma _{\alpha }(\varepsilon
)_{q^{\prime }\uparrow q\uparrow }=\Gamma _{\alpha \uparrow }$, $\Gamma
_{\alpha }(\varepsilon )_{q^{\prime }\downarrow q\downarrow }=\Gamma
_{\alpha \downarrow }$, and the polarization $P_{\alpha }=(\Gamma _{\alpha
\uparrow }-\Gamma _{\alpha \downarrow })/(\Gamma _{\alpha \uparrow }+\Gamma
_{\alpha \downarrow })$. If the two ferromagnets are made of the same
materials, then $P_{L}=P_{R}=P$, $\Gamma _{L\uparrow }=\Gamma _{R\uparrow
}\equiv \Gamma $, $\Gamma _{L\downarrow }=\Gamma _{R\downarrow }=\frac{1-P}{%
1+P}\Gamma $. We will take $I_{0}=e\Gamma /\hbar $ and $G_{0}=e^{2}/\hbar $
as scales, respectively, for the tunnel current and the differential
conductance, and hereafter take $\Gamma $ as an energy scale\cite{note1}. 
\begin{figure}[tb]
\begin{center}
\leavevmode\includegraphics[width=0.75\linewidth,clip]{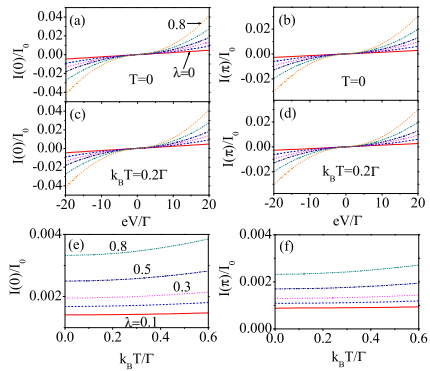}
\end{center}
\caption{(Color online) Tunnel current as a function of the bias voltage
(a)-(d) and of temperature (e)-(f) in parallel $I(0)$ and antiparallel $I(%
\protect\pi )$ configurations of magnetizations for different coupling
parameter $\protect\lambda =0,0.1,0.2,0.3,0.5,0.8$, where the polarization $%
P=0.5.$}
\label{fig2}
\end{figure}

The bias- and temperature-dependence of the tunnel current in the parallel
and antiparallel configurations of magnetizations are presented for
different coupling constant $\lambda $ of the MFL, as shown in Fig. 2. It is
seen that when $\lambda =0$, namely, the MFL recovers to the normal Fermi
liquid in this case, the tunnel current is proportional to the bias voltage
at small bias, suggesting that the system behaves an ohmic law in this case,
in agreement with the conventional result in the Fermi liquid. With
increasing the coupling constant $\lambda $, $I-V$ curves deviate obviously
the linear relation, and non-ohmic behaviors appear, i.e. the current
increases quadratically with the bias voltage. The larger the coupling $%
\lambda $, the more obvious the distinction from the ohmic behavior, as
illustrated in Figs. 2(a)-(d). This observation shows that the interactions
between electrons in the normal metal would have a remarkable effect on the
current-voltage characteristics where the Ohm law no longer holds. An
alternative reason for the nonlinearity of $I-V$ characteristics may be that
the energy dependent self-energy of the MFL in the central region leads to a
renormalization of the density of states which becomes energy dependent,
thereby resulting in a nonlinear voltage dependence of the current. When $%
\lambda $ is small, the tunnel current almost does not change with
temperature; while $\lambda $ becomes larger, the current increases slowly
with temperature, as shown in Figs. 2(e)-(f). This behavior also differs
from that in the usual Fermi liquid where the current decreases slowly with
increasing temperature, as thermal fluctuations enhance scatterings of
conduction electrons and thereby contribute to the resistance of the system.
It is interesting to note that the typical $I-V$ characteristics of Ni$_{80}$%
Fe$_{20}$/Co/Al-oxide junction (Figure 3.10 in Ref.\cite{miyazaki}) are very
similar to the shapes of the curves shown in our Figs. 2(a)-(d). 
\begin{figure}[tb]
\begin{center}
\leavevmode\includegraphics[width=0.75\linewidth,clip]{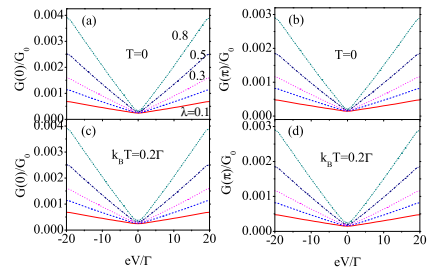}
\end{center}
\caption{(Color online) Differential conductance as a function of bias
voltage in parallel $G(0)$ and antiparallel $G(\protect\pi )$ configurations
for different coupling constant $\protect\lambda =0.1,0.2,0.3,0.5,0.8$ at
temperature $T=0$ (a)-(b) and $T=0.2\Gamma /k_{B}$ (c)-(d), where the
parameters are taken the same as in Fig. 1.}
\label{fig3}
\end{figure}

The differential conductance can be obtained by $G=dI(V)/dV$. The results
are shown in Figs. 3(a)-(d). As $\lambda =0$, the conductance is independent
of the bias voltage, which is nothing but the Ohm law. When $\lambda $ is
nonzero, the differential conductance behaves as $G=G_{0}+G_{1}V$ with $G_{0}
$ and $G_{1}$ nonzero constants at low biases. The non-ohmic behavior of $G$
comes from the interactions between conduction electrons via the exchange of
charge and spin fluctuations in the central region. The differential
conductance is observed to increase slowly with increasing temperature at
larger $\lambda $, and almost does not change when $\lambda $ is smaller
(e.g. $\lambda =0.1$). This observation is manifested itself in Figs.
2(e)-(f). We notice that the linear bias-dependence of the differential
conductance in various of junctions with La$_{1.85}$Sr$_{0.15}$CuO$_{4}$-In 
\cite{Kirtley} and even YBCO films \cite{Dagan} have also been observed. It
is worthy of noting that the differential conductance of a contact between
an ordinary metal and a MFL is shown to depend linearly on the applied
voltage\cite{Kupka}, where due to the asymmetry of electrodes, the
conductance for positive and negative biases is asymmetric. This result is
compatible with our observation. The origin of the linearity between the
conductance and the bias voltage could be explained by assuming charging
effects\cite{J.R.Kirtley}, the voltage-dependent tunneling penetration
probabilities\cite{R.Kirtley}, DOS effects \cite{Varma,Anderson}, inelastic
scattering\cite{Kirtley}, and so on. Our present study might offer a
different possibility, namely, such a linearity between $G$ and $V$ could
result from strong interactions between conduction electrons via exchanging
the charge and spin fluctuations. Since the real part of the self-energy
gives the correction of the single-particle energy, describing the elastic
scattering of quasiparticles, whereas the imaginary part determines the
lifetime of the quasiparticles, reflecting the inelastic scatterings.
Therefore, the linearity between $G$ and $V$ could also be dominated by the
inelastic scatterings between conduction electrons. 
\begin{figure}[tb]
\begin{center}
\leavevmode\includegraphics[width=0.75\linewidth,clip]{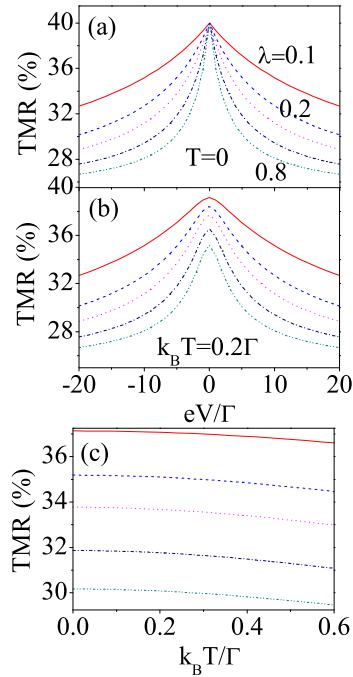}
\end{center}
\caption{(Color online) Tunnel magnetoresistance as a function of the bias
voltage as a function of bias voltage at $T=0$ (a) and $T=0.2\Gamma /k_{B}$
(b) and as a function of temperature (c) for different coupling constant $%
\protect\lambda =0.1,0.2,0.3,0.5,0.8$ at $V=5\Gamma /e$, where the
parameters are taken the same as in Fig. 1.}
\label{fig4}
\end{figure}

The TMR ratio as a function of the bias and temperature for different
coupling constant $\lambda $ is shown in Figs. 4(a)-(c). It is seen that the
TMR decreases with increasing the absolute magnitude of the bias, and is
symmetric to the zero-bias axis. The larger the coupling constant $\lambda $%
, the more rapidly decreasing the TMR, as presented in Figs. 4(a) and (b).
It suggests that the strong interactions between conduction electrons tend
to suppress the TMR ratio, which is a disadvantage for the application of
the FM-MFL-FM tunnel junction as a possible MRAM. This property of the TMR
has also been observed in various junctions (see Figure 3.7 in Ref.\cite%
{miyazaki}). One may observe that the TMR decreases slowly with increasing
temperature, as shown in Fig. 4(c).

The current and the TMR ratio as functions of the coupling constant $\lambda 
$ in the MFL for different temperatures are presented in Figs. 5(a)-(d). It
is found that the current depends linearly on the coupling constant $\lambda 
$ in the parallel or antiparallel alignment of magnetizations. This behavior
is also manifested in Figs. 2(a)-(d). It can be understood that, with the
increase of the coupling constant, the single-particle scattering rate which
is proportional to $\lambda $, increases, leading to that the quantum well
levels in the MFL could be broadened. Such a level broadening could make
more electrons tunnel through the barrier, thereby resulting in an increase
of the current with $\lambda $, as observed in Figs. 5(a) and (b). In either
case, $T=0$ or $T>0$, $I(0)$ is greater than $I(\pi )$, implying a spin
valve effect (see below). The TMR ratio is found to decay with increasing
the coupling constant $\lambda $, as shown in Figs. 5(c) and (d), suggesting
that the interactions of electrons are detrimental to the TMR effect. This
may be that the inelastic scatterings of electrons via exchanging the charge
and spin fluctuations weaken the spin-dependent scattering of electrons,
leading to that the TMR ratio decreases with increasing $\lambda $. 
\begin{figure}[tb]
\begin{center}
\leavevmode\includegraphics[width=0.75\linewidth,clip]{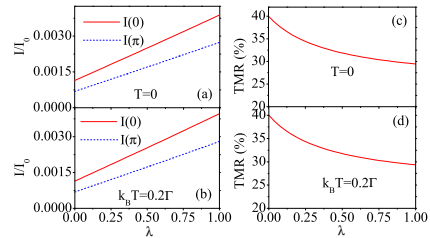}
\end{center}
\caption{(Color online) Tunnel current as a function of the coupling
constant $\protect\lambda $ at $T=0$ (a) and $T=0.2\Gamma /k_{B}$ (b);
Tunnel magnetoresistance as a function of the coupling constant $\protect%
\lambda $ at $T=0$ (c) and $T=0.2\Gamma /k_{B}$ (d), where $V=5\Gamma /e$,
and the other parameters are taken the same as in Fig. 1.}
\label{fig5}
\end{figure}

The relative angle $\theta $ dependences of the current as well as the TMR
ratio for different coupling constant are presented in Figs. 6(a)-(d). The
current as a function of $\theta $ shows a cosine-like shape, $G(\theta
)\sim \tilde{G}_{0}+\tilde{G}_{1}\cos \theta $ with $\tilde{G}_{0}$, $\tilde{%
G}_{1}$ constants, i.e., it decreases with increasing $\theta $ from zero to 
$\pi $, as illustrated in Figs. 6(a) and (b) for $T=0$ and $0.2\Gamma /k_{B}$%
, respectively. The TMR ratio as a function of $\theta $ shows a shape
similar to $(1-\cos \theta )$. These results display nothing but the
spin-valve effect. However, as discussed above, the coupling constant $%
\lambda $ tends to suppress the TMR effect.

\begin{figure}[tb]
\begin{center}
\leavevmode\includegraphics[width=0.75\linewidth,clip]{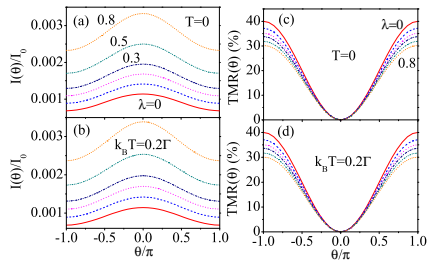}
\end{center}
\caption{(Color online) Tunnel current as a function of the relative
orientation angle $\protect\theta $ at $T=0$ (a) and $T=0.2\Gamma /k_{B}$
(b); Tunnel magnetoresistance as a function of the relative orientation
angle $\protect\theta $ at $T=0$ (c) and $T=0.2\Gamma /k_{B}$ (d), where the
coupling constant $\protect\lambda =0,0.1,0.2,0.3,0.5,0.8$, $V=5\Gamma /e$,
and the other parameters are taken the same as in Fig. 1.}
\label{fig6}
\end{figure}

In summary, we have discussed the spin-dependent transport in FM-MFL-FM
tunnel junctions. It is found that the current-voltage characteristics in
this system deviate obviously from the ohmic behavior, and the tunnel
current increases slightly with temperature, which are in contrast to those
of the system with a Fermi liquid where the Ohm law is satisfied. The TMR is
observed to decay exponentially with increasing the bias voltage, but to
decay slowly with increasing temperature. These results are qualitatively
consistent with the experimental observations found in various junctions,
suggesting that the present study might offer a possible different route to
understand the unusual experimental results of the $I-V$ and $G-V$
characteristics. With increasing the coupling constant of the MFL, the
current is shown to increase linearly, while the TMR is seen to decay
slowly. It appears that the interactions between electrons in the central
normal metal via exchanging the charge and spin fluctuations tend to
suppress the TMR effect. In addition, the spin-valve effect is also observed.

This work is supported in part by the National Science Foundation of China
(Grant Nos. 90103023, 10104015, 10247002), and by the Chinese Academy of
Sciences.

\end{document}